# Determination of Electron Extraction in Semiconductor Photoanodes: Steady-state and Small-perturbation Response

Paola Ragonese,[1] Chiara Maurizio[1], Boris Kalinic,[1] Thomas Kirchartz[2,3] and Sandheep Ravishankar[2*]

[1]Physics and Astronomy Department, University of Padova, via Marzolo 8, I-35131 Padova, Italy
[2]IMD-3 Photovoltaics, Forschungszentrum Jülich, 52425 Jülich, Germany
[3]Faculty of Engineering and CENIDE, University of Duisburg-Essen, Carl-Benz-Str. 199, 47057 Duisburg, Germany

*author for correspondence, email: s.ravi.shankar@fz-juelich.de

**Abstract**

This work develops an analytical model to consistently interpret the steady-state and small-perturbation response (both in the time and frequency domain) of photoanodes for solar water-splitting. In addition to accounting for the fundamental mechanisms of charge-carrier generation, recombination and slow hole transfer at the photoanode/electrolyte interface, the model overcomes the key shortcomings of existing models in the literature. These include consistency across measurements/bias conditions and the non-consideration of imperfect electron extraction at the collecting contact and its corresponding effect on the recombination rate in the bulk. We applied the model to analyse the time constants obtained from intensity-modulated photocurrent (IMPS) and photovoltage (IMVS) measurements of a hematite photoanode, obtaining an electron extraction velocity of 100 cm/s close to the 1 sun open-circuit potential, that corresponds to an electron mobility of 0.022 $cm^2V^{-1}s^{-1}$. The model further predicts a linear dependence of the photocurrent versus anodic voltage, an observation whose origin has been strongly debated in the literature in the case of hematite photoanodes. The generality of the model allows its extension to other photoanodes and photovoltaic systems, by the addition or removal of specific physical mechanisms.

**Introduction**

Photoelectrochemical (PEC) water oxidation is increasingly attracting attention due to its potential as an environmentally-friendly method for converting solar energy into hydrogen.[1] Since the Oxygen Evolution Reaction (OER) is considered the bottleneck of the overall process,[2] an efficient catalyst working as a photoanode is an important requirement. Although higher performances are currently obtained by coupling different materials through specific interface engineering,[3] the optimization of sustainable photoanodes made of a single material could offer a simpler solution. Hematite ($Fe_2O_3$) is a frequently studied material because it is earth abundant, low cost, stable and has a suitable band edge position with respect to the oxygen evolution reaction (OER), in addition to possessing an optimal energy band gap for a single-junction device (1.9-2.2 eV). However, its performance is limited by the high electron-hole recombination rate, poor electron conductivity and apparent short hole diffusion length.[4] Understanding the kinetic aspects of the involved processes, i.e. transport and recombination of photogenerated carriers in the bulk of the material and at the semiconductor-electrolyte interface is therefore crucial for the design of efficient photoanodes. However, despite continuous research endeavours, the multi-redox reaction at the photoanode-electrolyte interface involved in the OER still remains unclear.[5]

In this context, small-perturbation methods are powerful tools for probing the dynamic processes occurring in the photoanodes. In the time domain, these methods prominently include transient photocurrent (TPC) and transient photovoltage (TPV). In the frequency domain, the methods are impedance spectroscopy (IS), intensity-modulated photocurrent spectroscopy (IMPS) and intensity-modulated photovoltage spectroscopy (IMVS). In the case of IMPS, several models have been developed to distinguish the processes of hole charge transfer from the semiconductor to the electrolyte, trapping and recombination of charge carriers, and other multi-step reaction mechanisms.[6] TPC, which is the corresponding time-domain method of IMPS, has been used to estimate the rate of electron extraction and electron-hole recombination in the photoanode.[7] IS has been used to identify the existence of surface states in hematite photoanodes,[8] though their impact on the device performance has been extensively discussed in the literature.[9] IMVS measurements have been typically carried out at open-circuit conditions to determine parameters related to the recombination of charge carriers.[10] However, several of these models either do not fully capture the physics of operation of the photoanode (e.g. assuming perfect electron collection at the back contact) or lack self-consistency (e.g. cannot reproduce the steady-state response). Furthermore, it is unclear how these models can be unified to describe the response of the photoanode. Finally, since all these measurements are carried out on the same device, a unique equivalent circuit is required that generates the IS, IMPS and IMVS response.[11] Therefore, there is an urgent need to develop a consistent model that captures the basic processes occurring in the photoanode effectively and reproduces the steady-state and transient response (time-domain and frequency-domain).

In this work, we develop an optoelectronic model for semiconductor photoanodes to consistently interpret steady-state current-voltage measurements and the small-perturbation response in the frequency and time domain. In addition to the fundamental mechanisms of charge generation, recombination and slow hole kinetics at the photoanode/electrolyte interface, the model specifically accounts for the difference between the quasi-Fermi-levels of electrons and holes within the semiconductor absorber and the external applied voltage under illumination. Secondly, the model also accounts for non-ideal electron extraction at the collecting contact. Both these factors play a significant role in determining the current densities of recombination and charge transfer in the photoanode. The validity of the model was tested by comparing its predicted IMPS and IMVS spectra across different DC parameters with those obtained from full drift-diffusion simulations, which showed a good agreement with each other. Thus, our model can serve as a base model for predicting the steady-state and transient response of semiconductor photoanodes, to which specific physical models can be added (e.g. surface states) depending upon the system being studied.

Application of the model to the case of hematite photoanodes naturally explains the linear dependence of the measured photocurrent on the applied voltage, through a voltage-dependent electron extraction velocity that depends on the constant electric field across the hematite layer. Furthermore, the simulated IMPS and IMVS spectra show a good agreement with the trends of the experimental spectra reported in literature, as a function of DC parameters such as the applied voltage and the effect of addition of a hole scavenger. From a combined analysis of the time constants obtained from IMPS, IMVS and TPV measurements, we calculated a hole transfer rate constant of $k_{ct} = 16$ s$^{-1}$, confirming the slow kinetics of holes at the hematite/electrolyte interface. We further determined an electron extraction velocity of $S_{exc} = 100$ cm/s, that corresponds to an electron mobility of $\mu_n = 0.022$ cm$^2$V$^{-1}$s$^{-1}$ in the hematite layer.

**Results and Discussion**
For the sake of clarity, we begin by providing a brief description of the different time-domain (TPC, TPV) and frequency-domain methods (IMPS, IMVS) that are used in this work to study the physics of operation of our hematite photoanodes. The IMPS measurement involves the application of a small perturbation of the photon flux $\Phi$ at a chosen angular frequency $\omega$ on the sample, followed by measurement of the corresponding modulated current density $\tilde{\jmath}$. The IMPS transfer function is given by

$$Q(\omega) = \frac{\tilde{\jmath}}{\tilde{\jmath}_\Phi},  \quad (1)$$

where $\tilde{\jmath}_\Phi = q\tilde{\Phi}$ is the modulated photon flux represented as a current density. The IMPS transfer function is dimensionless and can be measured at any given DC condition of applied voltage or light intensity. The frequency $f$ ($\omega = 2\pi f$) is swept over a large range (typically mHz to kHz) to obtain the full IMPS spectrum. The corresponding measurement in the time domain is called a transient photocurrent (TPC) measurement. In a TPC measurement, the device is connected via a 50 Ω cable to a 50 Ω load, which corresponds to a situation close to short circuit for a typical device area of 0.1979 cm$^2$. A DC illumination intensity is then applied to the cell and the potential drop across the load is measured using an oscilloscope to obtain the DC photocurrent. A short laser pulse (with intensity much lower than the DC light intensity) impinges on the sample, which creates an excess photocurrent $\Delta j$ that rises and subsequently decays. Since TPC is a small-perturbation method, we assume that all processes are linearised with respect to the carrier concentration. For time constants that are well-separated in magnitude, the response can be analysed using[12]

$$\Delta j(t) = j_0 \left[ \exp\left(\frac{-t}{\tau_{j,\text{decay}}}\right) - \exp\left(\frac{-t}{\tau_{j,\text{rise}}}\right) \right], \quad (2)$$

where $j_0$ is a constant and $\tau_{j,\text{rise}}$, $\tau_{j,\text{decay}}$ are the characteristic time constants of the rise and decay of the photocurrent respectively.

The second method in the frequency domain is called intensity-modulated photovoltage spectroscopy (IMVS), which involves the measurement of the AC voltage $\tilde{V}$ upon application of an AC photon flux with a chosen angular frequency on the sample, for any given DC condition of applied voltage or light intensity. The IMVS transfer function is given by

$$W(\omega) = \frac{\tilde{V}}{\tilde{\jmath}_\Phi}, \quad (3)$$

which has the units of a resistance multiplied by area (Ωcm$^2$). We note that the output of IMPS or IMVS measurements is typically in the form of a responsivity, which can be easily converted to the desired form of the transfer functions shown in equations 1 and 3.[13] Similar to IMPS, the IMVS spectrum is measured over a large range of frequencies. The corresponding measurement in the time domain is called a transient photovoltage (TPV) measurement, which is identical to a TPC measurement except that the sample is held at open circuit by setting the load resistance to 1 MΩ. The evolution of the excess voltage $\Delta V$ versus time in a TPV measurement for the case of well-separated time constants in magnitude is given by[14]

$$\Delta V(t) = V_0 \left[ \exp\left(\frac{-t}{\tau_{V,\text{decay}}}\right) - \exp\left(\frac{-t}{\tau_{V,\text{rise}}}\right) \right], \quad (4)$$

where $V_0$ is a constant and $\tau_{V,\text{rise}}$, $\tau_{V,\text{decay}}$ are the characteristic time constants of the rise and decay of the phovoltage respectively.

The time constants in equations 2 and 4 are typically evolving as a function of the DC variables - light intensity, applied voltage and current density. Figure 1 shows the typical data representation of the small-perturbation frequency domain methods and their corresponding time domain methods. The frequency domain methods are expected to yield the same information as their corresponding methods in the time domain, since both domains are connected by the Laplace transform.[12a]

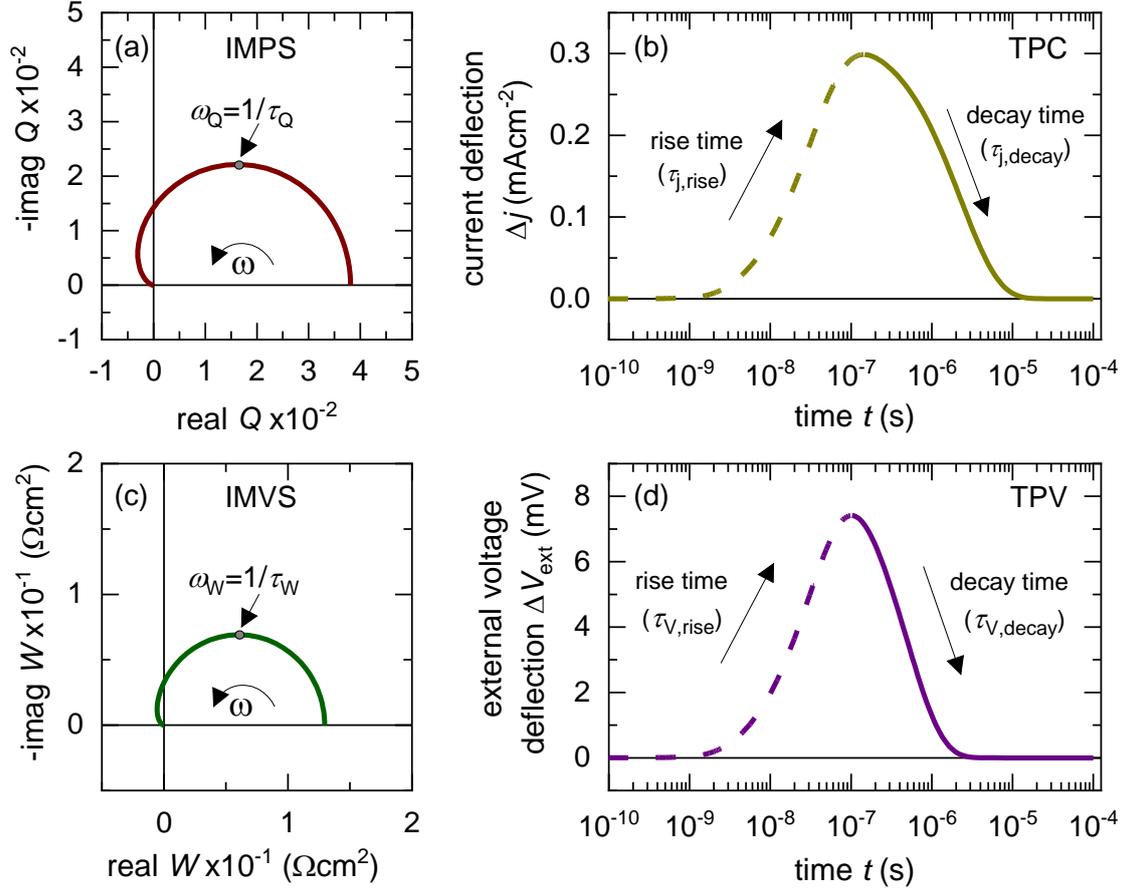

**Figure 1** Drift-diffusion simulations showing the typical representation of the spectra obtained from small-perturbation frequency domain measurements and their corresponding small-perturbation time domain measurements – (a) intensity-modulated photocurrent spectroscopy (IMPS) and (b) transient photocurrent (TPC) response, (c) intensity-modulated photovoltage spectroscopy (IMVS) and (d) transient photovoltage (TPV) response. In (b) and (d), the dashed lines indicate a rise of the quantity while solid lines indicate a decay. The characteristic time constants obtained from these spectra are indicated in the labels. Reproduced from ref.[12a]. © The authors of ref.[12a]. CC-BY 4.0.

*Two-capacitor model*

A fundamental model was developed by Wilson et al. to explain the transient response of a semiconductor-electrolyte junction after the application of a short light pulse.[15] The model assumes a doped semiconductor with a space-charge region whose capacitance (Fcm$^{-2}$) is $C_{sc}$, exchanging charge with the Helmholtz capacitor with capacitance $C_H$ at the photoanode-electrolyte interface. Minority carrier exchange between the capacitors at the photoanode/electrolyte interface is proportional to the difference of charge on the capacitors, with the respective rate constants (s$^{-1}$) of transfer given by $k_{sc}$ and $k_H$. The majority carriers are swept across the depletion region and collected, generating a current density $j$ in the external circuit. The model is completed by using Ohm's law which equates the sum of the potential drops across the space-charge and Helmholtz capacitors to the potential drop across the total series resistance $R_{tot}$ (sum of resistances arising from the electrolyte, semiconductor bulk and the Ohmic contact for electron collection) when the current density $j$ (Acm$^{-2}$) flows over it. This yields the equations

$$\frac{dQ_{sc}}{dt} = -j - k_{sc}(Q_{sc} - Q_H), \qquad (1)$$

$$\frac{dQ_H}{dt} = -j + k_H(Q_{sc} - Q_H), \tag{2}$$

$$\frac{Q_{sc}}{C_{sc}} + \frac{Q_H}{C_H} = jR_{tot}. \tag{3}$$

where $Q_{sc}$ and $Q_H$ are the charge per unit area (Ccm$^{-2}$) in the space-charge region and Helmholtz layer respectively. We first investigate the validity of the model at steady-state conditions. We therefore set the left-hand side of equations 1 and 2 to zero, which yields

$$\bar{j} = -k_{sc}(\bar{Q}_{sc} - \bar{Q}_H) = k_H(\bar{Q}_{sc} - \bar{Q}_H), \tag{4}$$

where the overbar indicates a steady-state value. Equation 4 implies

$$-k_{sc} = k_H, \tag{5}$$

that shows that the two rate constants are not independent of each other, rather their magnitudes are equal with opposite sign. The non-independency of the rate constants in this model is its fundamental limiting feature.

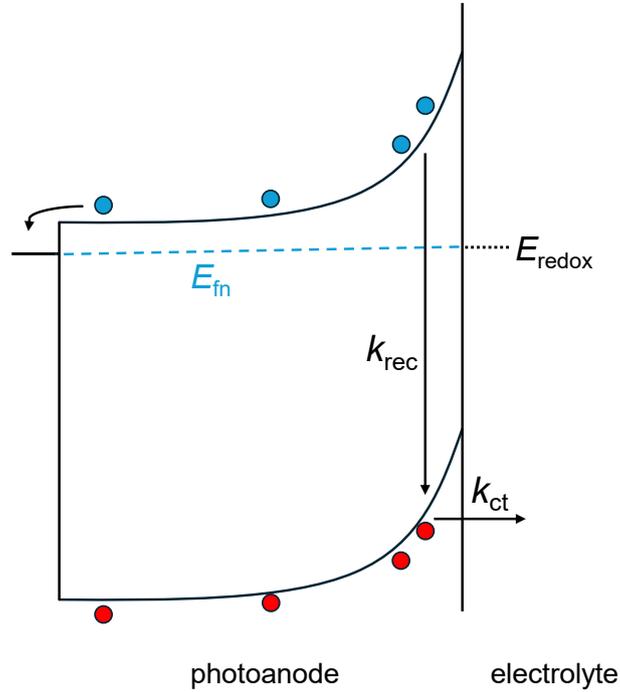

**Figure 2** Schematic of the processes considered in the modified two-capacitor model for IMPS measurements of photoanodes (ref.[6a]). The recombination of photogenerated electron-hole pairs at the photoanode/electrolyte interface is proportional to the rate constant $k_{rec}$ ($k_{sc}$ in the model of Wilson et al.[15]) and the transfer of holes from the valence band to the electrolyte is proportional to the rate constant $k_{ct}$ ($k_H$ in the model of Wilson et al.[15]). Perfect collection of electrons is assumed at the back contact.

Ponomarev and Peter modified this model to describe small-perturbation measurements of photoanodes in the frequency domain, specifically intensity-modulated photocurrent spectroscopy (IMPS) measurements.[6a] This was carried out by the addition of a photogenerated photon current density $j_\Phi = q\Phi$, where $\Phi$ is the photon flux, to equation 1. Furthermore, the rate constant $k_{sc}$ was replaced by $k_{rec}$, whose corresponding term now describes recombination of the photogenerated holes with the electrons in the conduction band. The rate constant $k_H$ was replaced by $k_{ct}$, indicating charge transfer of holes from the valence band to the electrolyte. A schematic of the processes considered in the model is shown in figure 2. Applying a small perturbation to equations 1-3, removing steady-state terms and using the Laplace transform, we obtain

$$i\omega \tilde{Q}_{sc} = -\tilde{j} + \tilde{j}_\Phi - k_{rec}(\tilde{Q}_{sc} - \tilde{Q}_H), \tag{6}$$

$$i\omega \tilde{Q}_H = -\tilde{j} + k_{ct}(\tilde{Q}_{sc} - \tilde{Q}_H), \tag{7}$$

$$\frac{\tilde{Q}_{sc}}{C_{sc}} + \frac{\tilde{Q}_H}{C_H} = \tilde{j} R_{tot}, \tag{8}$$

where the tilde indicates an AC quantity. We also define the following time constants

$$\tau_{ct} = \frac{1}{k_{ct}}, \tag{9}$$

$$\tau_{rec} = \frac{1}{k_{rec}}, \tag{10}$$

$$\tau_{par} = \left(\frac{1}{\tau_{ct}} + \frac{1}{\tau_{rec}}\right)^{-1} = \frac{1}{k_{ct}+k_{rec}}, \tag{11}$$

$$\tau_s = R_{tot}\left(\frac{1}{C_{sc}} + \frac{1}{C_H}\right)^{-1} = R_{tot} C_{tot}. \tag{12}$$

The IMPS transfer function can then be calculated from equations 6-8 as[11b]

$$Q = \frac{\tilde{j}}{\tilde{j}_\Phi} = Q_{ste} \frac{1+i\omega\gamma_c\tau_{ct}}{(1+i\omega\tau_s)(1+i\omega\tau_{par})}, \tag{13}$$

where $Q_{ste}$ is the low-frequency limit of the IMPS transfer function, which also corresponds to the charge transfer efficiency or surface transfer efficiency,[16] given by

$$Q_{ste} = \frac{\tau_{rec}}{\tau_{ct}+\tau_{rec}} = \frac{k_{ct}}{k_{ct}+k_{rec}}. \tag{14}$$

The variable $\gamma_c$ is a ratio of capacitances given by

$$\gamma_c = \frac{C_{tot}}{C_{sc}}. \tag{15}$$

The IMPS response of this model is shown in figure 3(a). We observe an arc in the upper quadrant at high frequencies that intercepts the real axis at the value $\gamma_c$ (termed the high-frequency intercept), followed by an arc in the lower quadrant that intercepts the real axis at the value $Q_{ste}$. The time constant of the high-frequency arc, obtained from the maxima $\omega_{HF}$ of the negative imaginary part of the transfer function versus angular frequency, is

$$\tau_{HF} = \frac{1}{\omega_{HF}} = \tau_s = R_{tot} C_{tot}. \tag{16}$$

The time constant of the low-frequency arc is given by

$$\tau_{LF} = \frac{1}{\omega_{LF}} = \tau_t = \frac{1}{k_{ct}+k_{rec}}. \tag{17}$$

For the case of fast charge transfer ($k_{ct} \gg k_{rec}$) shown in figure 3(b), the low-frequency arc moves to the upper quadrant and extends towards a value of $Q_{ste} = 1$. For the case of fast recombination ($k_{rec} \gg k_{ct}$) shown in figure 3(c), the low-frequency arc extends towards a value of $Q_{ste} = 0$. The spectrum in figure 3(a) offers a simple method to calculate the charge transfer and recombination time constants – using the time constant of the low-frequency arc (equation 17) and the low-frequency intercept on the real axis $Q_{ste}$ (equation 14). This method has been used widely in the photoelectrochemistry community to interpret intensity-modulated photocurrent spectroscopy (IMPS) measurements of photoanodes.[10a, 16-17] This model was extended by Ponomarev and Peter to include the effect of surface states[6a] and also multi-step charge transfer mechanisms.[6b]

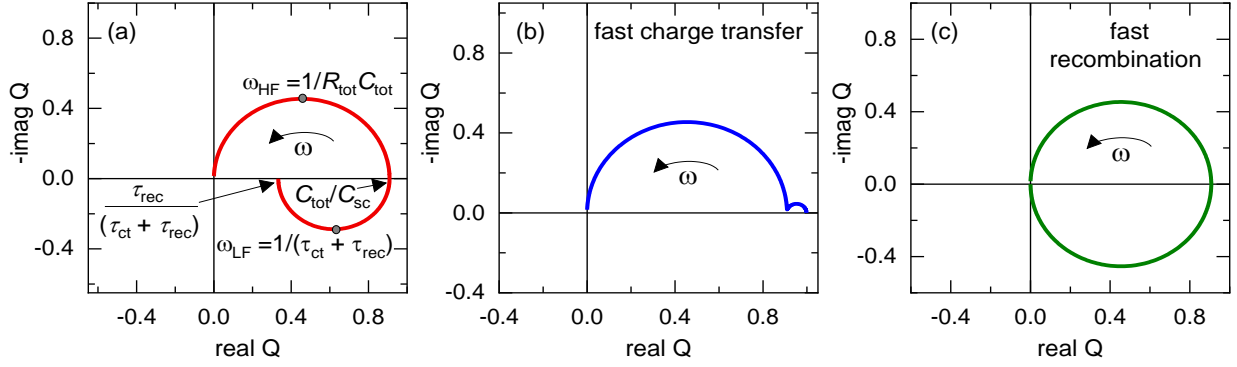

**Figure 3** (a) Simulated IMPS response of the two-capacitor model, developed in ref.[6a]. (b) corresponds to the IMPS spectrum for the case of fast charge transfer ($\tau_{ct} \ll \tau_{rec}$) while (c) corresponds to the spectrum for the case of fast recombination ($\tau_{rec} \ll \tau_{ct}$). The parameters used for the simulations are relative permittivity $\varepsilon_r = 32$, series resistance $R_s = 1$ $\Omega$cm$^2$, space charge capacitance $C_{sc} = 10^{-7}$ Fcm$^{-2}$, Helmholtz capacitance $C_H = 10^{-6}$ Fcm$^{-2}$. The hole transfer rate constant $k_{ct}$ was set to $k_{ct} = 10$ s$^{-1}$ in (a) and (c) and $k_{ct} = 10^4$ s$^{-1}$ in (b). The recombination rate constant $k_{rec}$ was set to $k_{rec} = 20$ s$^{-1}$ in (a) and (b) and $k_{rec} = 2 \times 10^4$ s$^{-1}$ in (c).

However, there are several problems with this model – (a) The steady-state solution of the model implies that $k_{ct}$ and $k_{rec}$ are not independent rate constants but are equal in magnitude (equation 5) and opposite in sign, leading to equations 11, 14 and 17 tending to infinity. (b) The intercept of the high-frequency arc on the real axis $\gamma_c$ is a ratio of capacitances, while it should actually depend on a ratio of resistances since the phase shift is zero for any point on the real axis.[18] (c) The model does not account for situations with an applied voltage, which prevents modelling spectra for different DC biases and also the corresponding IS/IMVS spectra. (d) The model does not account for the difference between the internal voltage (quasi-Fermi-level splitting) and the external (electrode) voltage under illumination (as seen in equation 8), discussed in detail in the next section. This difference is crucial to drive the photocurrent and to determine the voltage-dependent recombination losses in the absorber layer, which has been extensively discussed in the case of solar cells.[19] This also implies that equation 3, which equates the external potential to the sum of electrostatic potential drops in the photoanode, cannot be valid under illumination conditions. (e) The transport and collection of electrons (majority carriers) is assumed to be perfect, which is contrary to the observation of TPC decay time constants in the order of milliseconds.[7, 20] In this regard, a consistent model was developed by Bertoluzzi et al.,[11a] though they assume a simplified case where the perturbation of the electron concentration is introduced using a transport/series resistance.

*The internal voltage*

In a three-electrode measurement, the external voltage $V_{ext}$ is applied between the photoanode and the counter-electrode, while the measured voltage $V_{meas}$ is between the photoanode and a reference electrode. Therefore, $V_{meas}$ is connected to $V_{ext}$ through

$$V_{meas} = V_{OCP,dark} - V_{ext}, \tag{18}$$

where $V_{OCP,dark}$ is the measured dark open-circuit voltage (OCP). Therefore, a positive value of $V_{ext}$ (forward bias) makes $V_{meas}$ more negative, which corresponds to the cathodic region of the current-voltage curve, while a negative value of $V_{ext}$ (reverse bias) corresponds to the anodic region of the current-voltage curve.[21] We define the electrode voltage $V_{elec}$ as the difference between the quasi-Fermi-level of electrons at the collecting contact/photoanode interface minus the quasi-Fermi- level of holes at the photoanode/electrolyte interface, divided by the electronic charge $q$. Since a continuous quasi-Fermi-level is required to define a current, we have the hole quasi-Fermi-level at the photoanode/electrolyte interface ($x = d$) equal to the redox level $E_{\text{redox}}$ for the OER. For a current density $j$ flowing through the external circuit, $V_{\text{elec}}$ is connected to the external voltage $V_{\text{ext}}$ through

$$V_{\text{elec}} = V_{\text{ext}} + R_s j, \qquad (19)$$

where $R_s$ is the external series resistance that includes the potential drop in the electrolyte from the counter-electrode to the redox Fermi level at the photoanode/electrolyte interface. A positive value of $j$ implies a photocurrent flowing out of the photoanode with $V_{\text{elec}} > V_{\text{ext}}$, while a negative value implies a current flowing into the photoanode with $V_{\text{elec}} < V_{\text{ext}}$.

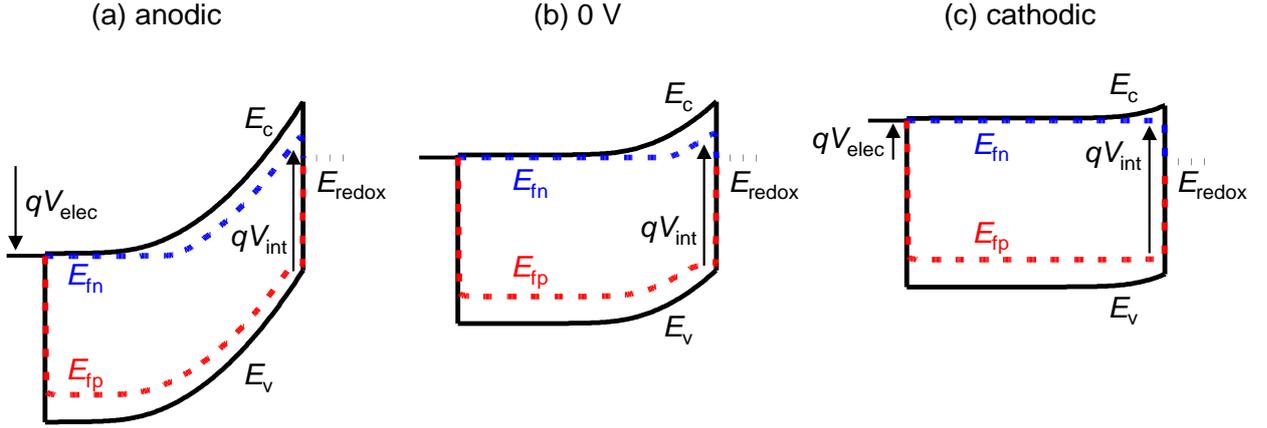

**Figure 4** Drift-diffusion simulations of the band diagrams of an idealised (highly doped, n-type) hematite photoanode under 1 sun illumination at (a) +1.23 V (anodic) (b) 0 V and (c) -0.51 V (cathodic). The slow kinetics of hole transfer from the photoanode to the electrolyte leads to accumulation of holes at the photoanode/electrolyte interface and causes a sharp jump of the hole quasi-Fermi-level to the redox level. Simulation parameters are shown in table S1 in the SI.

To explain this effect in further detail, we carry out drift-diffusion simulations of an idealised (i.e. highly doped) hematite photoanode, consisting of a hematite absorber layer and a fluorine-doped tin oxide (FTO) layer, that acts as the contact for electron collection. The drift-diffusion simulations are explained in detail in section A1 in the SI and simulation parameters are shown in table S1 in the SI. Figure 4 shows simulated band diagrams of the FTO/hematite/electrolyte junction under 1 sun illumination, at different applied voltages. At 0 V under illumination, we observe a depletion region at the hematite/electrolyte interface due to the high dopant (donor) density in the $Fe_2O_3$ layer, followed by a neutral region in the bulk. The width of the depletion region is enlarged in the case of anodic voltages while it is reduced in case of cathodic voltages. In all cases, we observe that the hole quasi-Fermi-level is very close to the valence band at the $Fe_2O_3$/electrolyte interface. This can be understood by considering that the rate of holes flowing from the valence band to the electrolyte is typically described by

$$\frac{dp}{dt} = k_{\text{ct}}(p - p_0), \qquad (20)$$

where $p$ is the hole density at the hematite/electrolyte interface, $p_0$ is the corresponding equilibrium hole density and $k_{\text{ct}}$ is a rate constant for hole transfer. Since the kinetics of hole transfer at the hematite/electrolyte interface is well-known to be very slow,[7, 22] a large density of holes is required to drive a given current density. The hole quasi-Fermi-level thus remains close to the valence band at the interface, followed by a sharp step down to the redox level.

Furthermore, in all cases in figure 4, a large internal voltage $V_{int}$ exists in the hematite layer, where $V_{int}$ is proportional to the average quasi-Fermi-level splitting $\Delta E_f$ in the layer i.e. $V_{int} = \Delta E_f/q$. Thus, we conclude that in the range of applied voltage where the photocurrent is much higher than the dark current, i.e., the typical range of polarization of photoanodes, $V_{int}$ is much larger than $V_{elec}$. In the case of solar cells, this difference between $V_{int}$ and $V_{elec}$ has been shown to have a significant impact on the efficacy of charge extraction.[12b, 14, 19a, 19b] Figure 4 shows that $V_{int}$ is much larger than the $V_{elec}$ for typical operation conditions of the photoanode, which implies significant recombination currents at each point of the current-voltage curve because the recombination rate $U \propto \exp(\Delta E_f/k_B T) = \exp(qV_{int}/k_B T)$. In the case of solar cells, this effect significantly contributes to the differences between the recombination current in the dark and under illumination.[19a, 19c] In summary, the internal voltage is an important factor that controls the rates of charge extraction and recombination, which implies that the assumption of $V_{int} = V_{elec}$ cannot be made while modelling the optoelectronic response of the photoanode.

*Electron exchange model*

We will now develop a model that overcomes the shortcomings of the two-capacitor model and hence captures the physics of operation of the photoanode/electrolyte junction effectively. We use a hematite photoanode as our reference device, whose fabrication details are reported in section A3 in the SI. The current-voltage curves of the hematite photoanode were recorded under AM 1.5 G illumination and in the dark condition, in 1M KOH (figure 5(a)). The dark current is minimal at low potential up to 1.65 V, when the OER occurs.

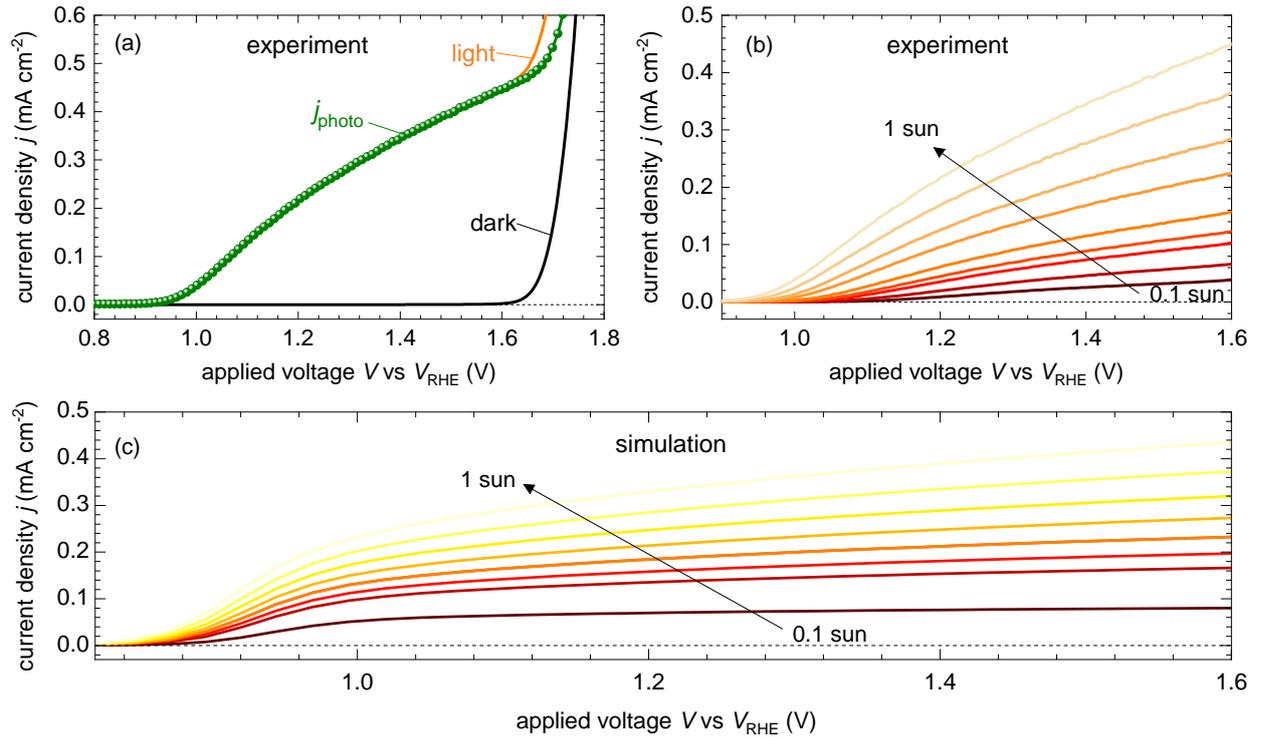

**Figure 5** (a) Experimental current-voltage curves of a hematite photoanode in KOH 1M solution, in the dark and under AM 1.5G illumination (front). Also shown is the photocurrent versus voltage, which is defined as $j_{photo} = j_{light} - j_{dark}$. (b) Corresponding current-voltage curves for different light intensities. (c) Simulated current-voltages curves for different light intensities using the developed analytical model. Simulation parameters are shown in table S2 in the SI. For (c), the built-in voltage was set to 0.1 V, the hole transfer rate constant was set to $k_{ct} = 10$ s$^{-1}$ and the recombination coefficient (defined in equation 24) was set to $B_{rec} = 10^{-11}$ cm$^3$s$^{-1}$.

Figure 5(a) shows the slope of the photocurrent $j_{photo}$ (calculated as the difference between the light and dark current-voltage curves) from its onset at 0.95 V up to 1.65 V (beyond which the steep rise is due to the dark current). Figure 5(b) shows that as the intensity of the light increases, the resulting photocurrent also rises while maintaining a mostly linear dependence on voltage. This shape of the voltage-dependent photocurrent can arise due to different mechanisms, such as the shunt resistance,[23] charge collection dominated by the depletion layer (Gärtner model[24]) or the drift of charge carriers across an electric field in the bulk of the photoanode.[4] In this regard, Klahr et al.[23] ruled out the shunt resistance and the depletion layer-dominated charge collection (which would imply that the photocurrent is proportional to the square root of the applied anodic voltage) as the originating mechanism. They instead attributed this effect to the drift of holes due to the built-in electrostatic voltage $V_{bi}$ across the bulk of the photoanode.[23] The argument of an electrostatic potential across the entire photoanode is consistent with our previous work, where we clarified that the large doping densities observed from capacitance measurements of hematite photoanodes are an artefact arising due to the response of the geometric capacitance.[21] This implies that the effective doping densities are low enough for the electrode charge to determine the potential distribution, which creates a constant electric field (linear electrostatic potential drop) in the photoanode, as shown in the simulated band diagrams under illumination in figure 6. The slow hole kinetics at the hematite/electrolyte interface leads to the accumulation of holes at this interface that causes the hole Fermi level to approach the valence band closely before making a step down to the redox Fermi level of the electrolyte, as discussed previously. The electron density remains relatively constant through the bulk of the hematite layer, increasing significantly close to the collecting contact while dropping to the equilibrium concentration at the photoanode/electrolyte interface. The existence of a constant electric field through the absorber leads to a voltage-dependent charge extraction of electrons from the bulk to the collecting contact (and hence a voltage-dependent photocurrent), which is discussed in detail in subsequent sections.

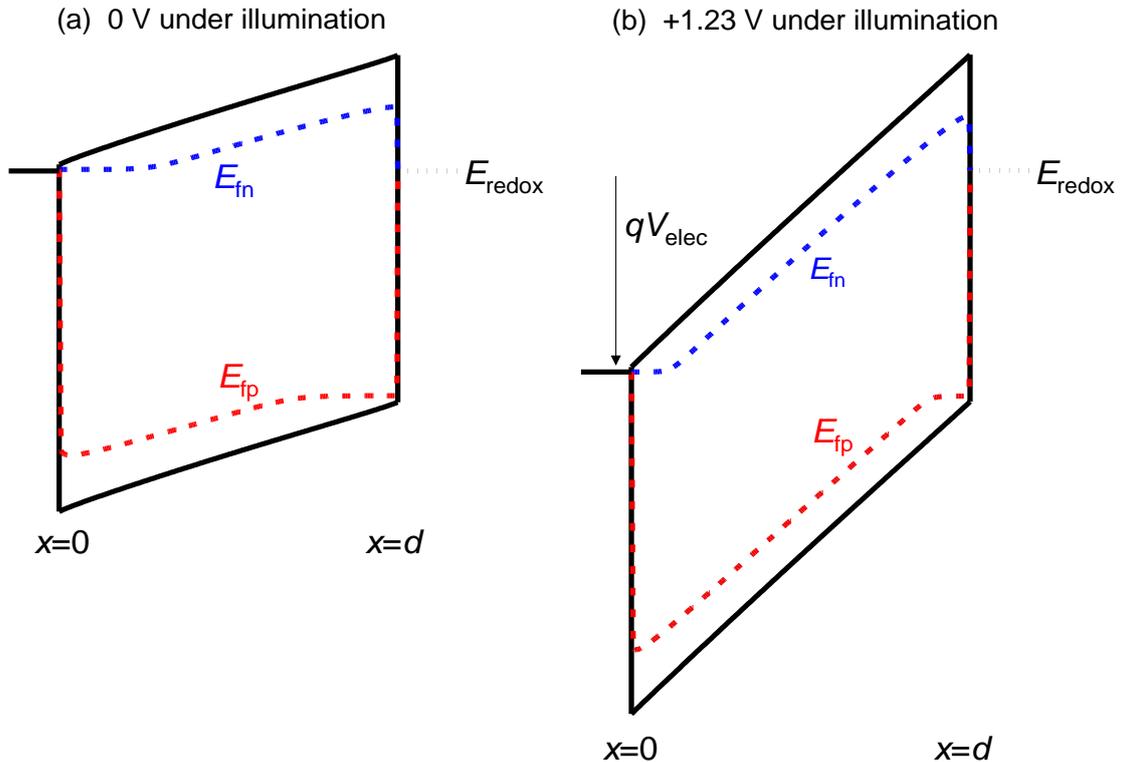

**Figure 6** Drift-diffusion simulations of the band diagrams of an intrinsic (zero doping density) hematite photoanode under 1 sun illumination at (a) 0 V and (b) +1.23 V. $V_{\text{elec}}$ is the electrode voltage, defined in equation 19. Simulation parameters are shown in table S1 in the SI. The interface layer mobilities of electrons and holes was increased by a factor of 10 from the value in table S1, which is an arbitrarily low value to simulate the effect of the slow hole transfer at the photoanode/electrolyte interface (see section A1 in the SI).

The model considers a photoanode of thickness $d$ with a density of positively-charged dopants $N_{\text{d}}$ distributed uniformly. Since the potential distribution is assumed to be determined by the electrode charge and not by the dopants, a constant electric field $F$ is assumed through the photoanode, which leads to a built-in voltage $V_{\text{bi}}$ at the photoanode/electrolyte interface ($x = d$) (schematic shown in section A4 in the SI). The electrons are collected by the contact at $x = 0$ and the holes are extracted to the electrolyte at $x = d$. We further assume that the electron contact is perfectly blocking for holes and the photoanode/electrolyte interface is perfectly blocking for electrons (i.e. $j_{\text{p}}(0) = 0$ and $j_{\text{n}}(d) = 0$). We also assume uniform photo-generation across the absorber layer. Under a steady-state condition of large anodic applied voltage and one sun illumination, we have a density of electrons $n_{\text{s}}$ and a density of holes $p_{\text{s}}$ accumulate at the photoanode/electrolyte interface in a layer of thickness $d_{\text{int}}$. The current density of holes transferred to the electrolyte is given by

$$j_{\text{p}}(d) = q d_{\text{int}} k_{\text{ct}} (p_{\text{s}} - p_{\text{s0}}), \tag{21}$$

where $k_{\text{ct}}$ is the rate constant for hole transfer and $p_{\text{s0}}$ is the equilibrium hole density at the photoanode/electrolyte interface. The electrons are transported across the bulk of the photoanode to the collecting contact at $x = 0$. Under the assumption of an exponentially-decaying electron current density across the photoanode (which is a valid assumption for low current densities and significant bulk recombination, see figure S3 in the SI), the measured current density of electrons at the collecting contact is given by (see section A4 in the SI for the derivation)

$$j_{\text{n}}(0) = q S_{\text{exc}} \left[ n_{\text{s}} - n_{\text{s0}} \exp\left(\frac{q V_{\text{elec}}}{k_{\text{B}} T}\right) \right]. \tag{22}$$

where $n_{\text{s0}}$ is the equilibrium electron concentration at the photoanode/electrolyte interface. $S_{\text{exc}}$ is the electron extraction velocity for the photoanode, given by

$$S_{\text{exc}} = \frac{\mu_{\text{n}} F \left(1 - \frac{k_{\text{B}} T}{q F d}\right)}{\left[\exp(-1) - \exp\left(\frac{-q F d}{k_{\text{B}} T}\right)\right]}, \tag{23}$$

where $\mu_{\text{n}}$ is the electron mobility and $F = (V_{\text{bi}} - V_{\text{elec}})/d$ is the electric field, with $V_{\text{bi}}$ being the built-in electrostatic potential. Furthermore, we assume that the recombination at the photoanode/electrolyte interface is governed by the product of $n_{\text{s}}$ and $p_{\text{s}}$. For clarity, we replace $j_{\text{n}}(0)$ with $j$, the measured current density. The main advantage of our approach is the simplification of the problem by requiring only the solution of the charge densities at the photoanode/electrolyte interface to determine the total measured current density. We obtain the continuity equations for the electron and hole density at the photoanode/electrolyte interface as

$$\frac{dn_{\text{s}}}{dt} = \frac{-j}{q d_{\text{int}}} + \frac{j_{\Phi}}{q d} - B_{\text{rec}} (n_{\text{s}} p_{\text{s}} - n_{\text{i}}^2), \tag{24}$$

$$\frac{dp_{\text{s}}}{dt} = \frac{j_{\Phi}}{q d} - B_{\text{rec}} (n_{\text{s}} p_{\text{s}} - n_{\text{i}}^2) - k_{\text{ct}} (p_{\text{s}} - p_{\text{s0}}), \tag{25}$$

where $j_{\Phi}$ is the photogenerated current density represented as a current density ($j_{\Phi} = \beta \times q\Phi$, where $\Phi$ is the input photon flux and $\beta$ is the efficiency of charge carrier photogeneration, assumed to be equal to 1 for simplicity) and $B_{\text{rec}}$ is the recombination coefficient (cm³/s). The internal voltage (quasi-Fermi-level splitting) is related to the densities of electrons and holes via

$$n_s p_s = n_i^2 \exp\left(\frac{qV_{\text{int}}}{k_B T}\right), \tag{26}$$

where $n_i = \sqrt{n_{s0} p_{s0}} = \sqrt{N_c N_v} \exp(-E_g/2k_B T)$ is the intrinsic carrier concentration, $n_{s0}$ and $p_{s0}$ are the equilibrium electron and hole densities at the photoanode/electrolyte interface, $N_c$ and $N_v$ are the effective density of states in the conduction and valence band respectively and $E_g$ is the bandgap. The equilibrium hole density at the photoanode/electrolyte interface $p_{s0}$ was determined from the equation for $n_i$ after calculating $n_{s0}$ using $n_{s0} = N_c \exp(-(E_{bn} + qV_{bi})/k_B T)$, where $E_{bn}$ is the injection barrier for electrons at the FTO/hematite interface. At steady-state, using equations 22, 24 and 25, we have

$$j = qS_{\text{exc}}\left[n_s - n_{s0}\exp\left(\frac{qV_{\text{elec}}}{k_B T}\right)\right] = q d_{\text{int}} k_{\text{ct}} (p_s - p_{s0}). \tag{27}$$

The steady-state solutions of $n_s$ and $p_s$ are shown in section A5 in the SI, which allows determination of the current density from equation 27. Figure 5(c) shows simulated current-voltage curves as a function of light intensity using the analytical model, showing similar features as the measured current-voltage curves in figure 5(b) (we note that the model does not encompass the dark photocurrent at large anodic potentials observed in figure 5(a)). Figure 7 shows the effect of the hole transfer rate constant $k_{\text{ct}}$, recombination coefficient $B_{\text{rec}}$, electron extraction velocity $S_{\text{exc}}$ and the electrostatic built-in potential $V_{\text{bi}}$ on the simulated steady-state current-voltage curves. These curves show the characteristic 'S'- shape typical of hematite photoanodes, with a photocurrent that depends linearly on the applied voltage (at high anodic voltages) due to the proportionality between $S_{\text{exc}}$ and the electric field $F$ (for large anodic voltages and hence large electric fields, $S_{\text{exc}} \propto \mu_n F$, see equation 23). The shape of the current-voltage curve is strongly influenced by the slow kinetics of the holes at the photoanode/electrolyte interface, which leads to a large density of holes accumulating at this interface and a large $V_{\text{int}}$ through almost the entire current-voltage curve (charge densities and internal voltage shown in figure S4 in the SI), which leads to large recombination current densities. An increase in $k_{\text{ct}}$ reduces this hole accumulation, which results in higher current densities and shifts the photocurrent onset toward more cathodic potentials (figure 7(a)). Experimentally, similar effects are observed by the addition of a hole scavenger (e.g. $H_2O_2$).[25] Conversely, an increase in the recombination coefficient leads to lower current densities and slightly shifts the onset potential to more anodic values (figure 7(b)). This behaviour aligns with the observed performance of optimized hematite photoanodes, where improved charge transfer efficiency is achieved through reduced interfacial electron-hole recombination.[26] Figure 7(c) shows the effect of $S_{\text{exc}}$ by varying the electron mobility $\mu_n$. Lower values of $S_{\text{exc}}$ increase the $V_{\text{int}}$ (see figure S5 in the SI), leading to higher recombination current density $j_{\text{rec}}$ since $j_{\text{rec}} \propto \exp(qV_{\text{int}}/n_{\text{id}} k_B T)$, where $n_{\text{id}}$ is the ideality factor.[23] Higher values of $S_{\text{exc}}$ increase the photocurrent by improving the efficiency of electron extraction to the collecting contact, though it does not have a significant impact on the photocurrent onset potential. This is in agreement with the observed increase in photocurrent when electron extraction to the FTO is improved, by engineering the FTO/hematite interface with an ITO or $TiO_2$ underlayer.[27] Finally, the effect of the $V_{\text{bi}}$ is shown in figure 7(d). Increasing the $V_{\text{bi}}$ creates a cathodic shift in the onset of the photocurrent density and also leads to an increased photocurrent. This is because $S_{\text{exc}}$ increases with increasing $V_{\text{bi}}$ (equation 23), which leads to improved electron collection.

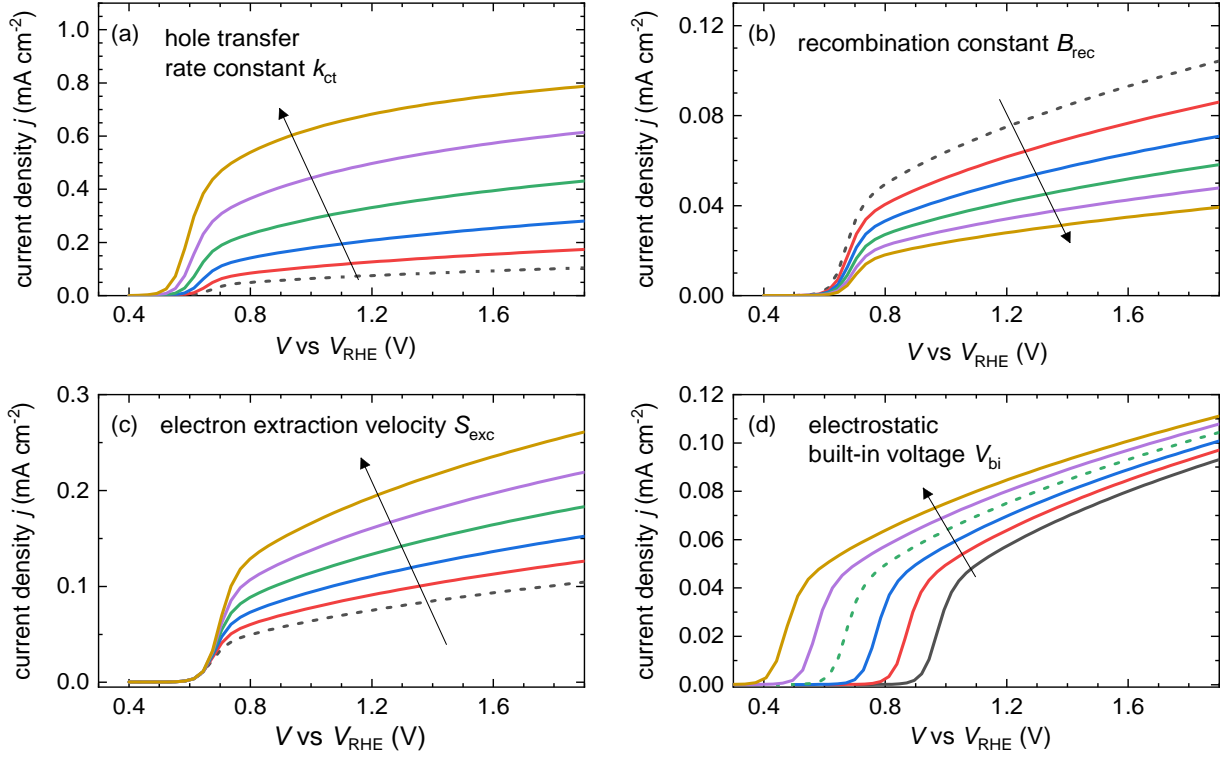

**Figure 7** Simulated current-voltage curves using our analytical model for different values of (a) hole transfer rate constant $k_{ct}$ (1-243 s$^{-1}$), (b) recombination constant $B_{rec}$ (5×10$^{-11}$ – 3.8×10$^{-10}$ cm$^3$/s), (c) electron mobility $\mu_n$ (0.01– 0.0759 cm$^2$V$^{-1}$s$^{-1}$) that modifies the electron extraction velocity $S_{exc}$ (see equation 23) and (d) electrostatic built-in voltage $V_{bi}$ (0.1-0.6 V). See table S2 in the SI for all the simulation parameters. The applied voltage versus $V_{RHE}$ is the applied potential minus the measured OCP measured in the dark condition (from data in figure 5). The dashed lines are obtained using the parameters in table S2 in the SI, used as reference in each graph.

*Small-perturbation response*

We now develop the model for the different small-perturbation methods commonly used, namely intensity-modulated photovoltage spectroscopy (IMVS), intensity-modulated photocurrent spectroscopy (IMPS), impedance spectroscopy (IS) in the frequency domain, and transient photovoltage (TPV) and transient photocurrent (TPC) in the time domain. Applying a small perturbation to equations 24, 25 and removing steady-state terms, we obtain

$$\frac{d\tilde{n}_s}{dt} = \frac{-\tilde{j}}{qd_{int}} + \frac{\tilde{j}_\Phi}{qd} - B_{rec}p_s\tilde{n}_s - B_{rec}n_s\tilde{p}_s, \tag{28}$$

$$\frac{d\tilde{p}_s}{dt} = \frac{\tilde{j}_\Phi}{qd} - B_{rec}p_s\tilde{n}_s - B_{rec}n_s\tilde{p}_s - k_{ct}\tilde{p}_s, \tag{29}$$

where the tilde indicates a modulated quantity. The modulated electron current density is obtained by applying a small perturbation to equation 22, yielding

$$\tilde{j} = qS_{exc}\tilde{n}_s - \frac{\tilde{V}_{elec}}{R_{exc}}, \tag{30}$$

where $R_{exc}$ is the resistance associated with the exchange of electrons from the photoanode to the collecting electrode, given by

$$R_{exc} = \left(\frac{k_BT}{q^2 S_{exc}n_{s0}}\right)\exp\left(\frac{-qV_{elec}}{k_BT}\right). \tag{31}$$

Substituting equation 30 in equation 28, we obtain

$$\frac{d\tilde{n}_s}{dt} = \frac{-\tilde{n}_s}{\tau_{exc}} + \frac{\tilde{V}_{elec}}{qd_{int}R_{exc}} + \frac{\tilde{J}_\Phi}{qd} - B_{rec}p_s\tilde{n}_s - B_{rec}n_s\tilde{p}_s , \qquad (32)$$

where $\tau_{exc}$ is a time constant for electron extraction from the photoanode/electrolyte interface, given by[12b]

$$\tau_{exc} = \frac{d_{int}}{S_{exc}} . \qquad (33)$$

We further define the recombination lifetime of the electrons as

$$\tau_n = \frac{1}{B_{rec}p_s} \qquad (34)$$

and the recombination lifetime of the holes as

$$\tau_p = \frac{1}{B_{rec}n_s} . \qquad (35)$$

We can rewrite equation 32 using equations 34-35 as

$$\frac{d\tilde{n}_s}{dt} = \frac{-\tilde{n}_s}{\tau_{exc}} + \frac{\tilde{V}_{elec}}{qd_{int}R_{exc}} + \frac{\tilde{J}_\Phi}{qd} - \frac{\tilde{n}_s}{\tau_n} - \frac{\tilde{p}_s}{\tau_p} . \qquad (36)$$

In a transient measurement, the total current is given by the balance of the extracted electrons and the capacitive discharge from the electrode. In addition, we also consider a current loss through an effective shunt resistance $R_{sh}$, which yields

$$\tilde{J}_{tot} = \tilde{J} - C_g \frac{d\tilde{V}_{elec}}{dt} - \frac{\tilde{V}_{elec}}{R_{sh}} . \qquad (37)$$

The model is completed by the relation between $V_{elec}$ and the applied external voltage $V_{ext}$ through the series resistance $R_s$, which yields

$$\tilde{V}_{elec} = \tilde{V}_{ext} + R_s\tilde{J}_{tot}. \qquad (38)$$

Equations 29, 36, 37 and 38 can be arranged into a 3x3 matrix and solved (see section A6 in the SI), yielding the IMPS transfer function as

$$Q(\omega) = Q_1\left(\frac{\tau_{j,1}}{1+i\omega\tau_{j,1}}\right) + Q_2\left(\frac{\tau_{j,2}}{1+i\omega\tau_{j,2}}\right) + Q_3\left(\frac{\tau_{j,3}}{1+i\omega\tau_{j,3}}\right), \qquad (39)$$

where $Q_1$, $Q_2$ and $Q_3$ are constants and $\tau_{j,1}$, $\tau_{j,2}$ and $\tau_{j,3}$ are characteristic time constants. Similarly, the IMVS transfer function can be obtained from the solution of the IMVS matrix (see section A6 in the SI) as

$$W(\omega) = W_1\left(\frac{\tau_{V,1}}{1+i\omega\tau_{V,1}}\right) + W_2\left(\frac{\tau_{V,2}}{1+i\omega\tau_{V,2}}\right) + W_3\left(\frac{\tau_{V,3}}{1+i\omega\tau_{V,3}}\right), \qquad (40)$$

where $W_1$, $W_2$ and $W_3$ are constants and $\tau_{V,1}$, $\tau_{V,2}$ and $\tau_{V,3}$ are characteristic time constants. The time constants in equations 39 and 40 are obtained from the inverse eigenvalues of the IMPS and IMVS matrices shown in section A6 in the SI. A simple analytical formula for these time constants is unfortunately impossible due to the complicated coupling between the electron transfer, hole transfer and recombination processes.

Figure 8 shows the simulated IMPS and IMVS spectra as a function of different DC parameters, for the specific case of slow hole transfer to the electrolyte (i.e. $\tau_{ct} \gg \tau_{exc}$). In general, we observe two arcs for both IMPS and IMVS, one in the upper quadrant at high frequencies followed by an arc in the lower quadrant at low frequencies. Experimental measurements also typically show two arcs for both IMPS and IMVS,[10a, 16, 28] though Klotz et al. observed the LF arc in the upper quadrant in the case of IMVS.[28] The effect of the DC photon flux or light intensity on the IMPS and IMVS spectra is shown in figures 8(a) and 8(b) respectively. Increasing the photon flux causes the width of both the arcs to decrease and leads to smaller values of the intercepts on the real axis. In the case of IMPS, the intercepts on the real axis can be interpreted as ratios of resistances,[18b] with the low-frequency (LF) intercept corresponding to the differential quantum efficiency[18a] (termed $EQE_{PV-diff}$) or the charge transfer efficiency.[11b] The reduction of the intercept values thus implies an increased recombination rate (reduced recombination resistance) upon generation of electron-hole pairs with increasing light intensity, exacerbated by the slow extraction of the holes. In the case of IMVS, the widths of the arcs correspond to the magnitudes of different resistances and hence, the reduction of the arc widths with increasing light intensity can be viewed as the corresponding reduction of the magnitudes of the resistances.

The effect of the hole transfer rate constant $k_{ct}$ is shown in figures 8(c) and 8(d). In the case of IMPS, increasing $k_{ct}$ leads to larger magnitudes of the high-frequency (HF) and LF intercepts (higher charge transfer efficiency). This can be understood as the reduction of the interfacial hole density due to faster hole transfer to the electrolyte, which reduces the recombination rate and allows more electrons to be collected by the electrode, leading to a larger current density. In the limiting case of very large $k_{ct}$, the LF arc disappears and the HF intercept approaches the maximum value of 1. Such a trend in the IMPS spectra is expected to occur, for example, upon the addition of a hole scavenger and has indeed been observed experimentally.[16, 28] A similar effect is observed in the IMVS spectra, with the LF intercept increasing in magnitude for higher $k_{ct}$ until the LF arc disappears. This behaviour has also been observed from IMVS measurements of hematite photoanodes using a hole scavenger.[10b, 16] For very high $k_{ct}$ values, the LF arc moves to the upper quadrant.

The effect of the recombination coefficient $B_{rec}$ on the IMPS and IMVS spectra is shown in figures 8(e) and 8(f). Increasing values of $B_{rec}$ lead to a reduction in the magnitude of the HF and LF intercepts on the real axis for both these techniques. In the case of IMVS, the concomitant reduction in the width of the HF arc implies a reduced HF resistance associated with increased recombination.

The influence of the charge extraction velocity $S_{exc}$ on the IMPS and IMVS spectra is shown in figures 8(g) and 8(h). In the case of IMPS, we observe a HF arc in the upper quadrant and a LF arc in the lower quadrant, whose intercepts on the real axis increase in magnitude with increasing $S_{exc}$. For sufficiently large $S_{exc}$ values, the LF arc disappears, similar to the effect of $k_{ct}$. In the case of IMVS, an arc each in the upper and lower quadrant is observed (for higher $k_{ct}$ values, the LF arc is shifted to the upper quadrant, shown in figure 8(d)), which are mostly unaltered at lower frequencies for different $S_{exc}$. However, at high frequencies, the real part of the transfer function makes a transition to negative values before reaching the origin. This transition is also observed for the IMPS spectra and is pronounced for lower values of $S_{exc}$. This spectral feature is thus a signature of non-ideal electron extraction and can be used to estimate the magnitude of $S_{exc}$ experimentally, which is discussed in the next section.

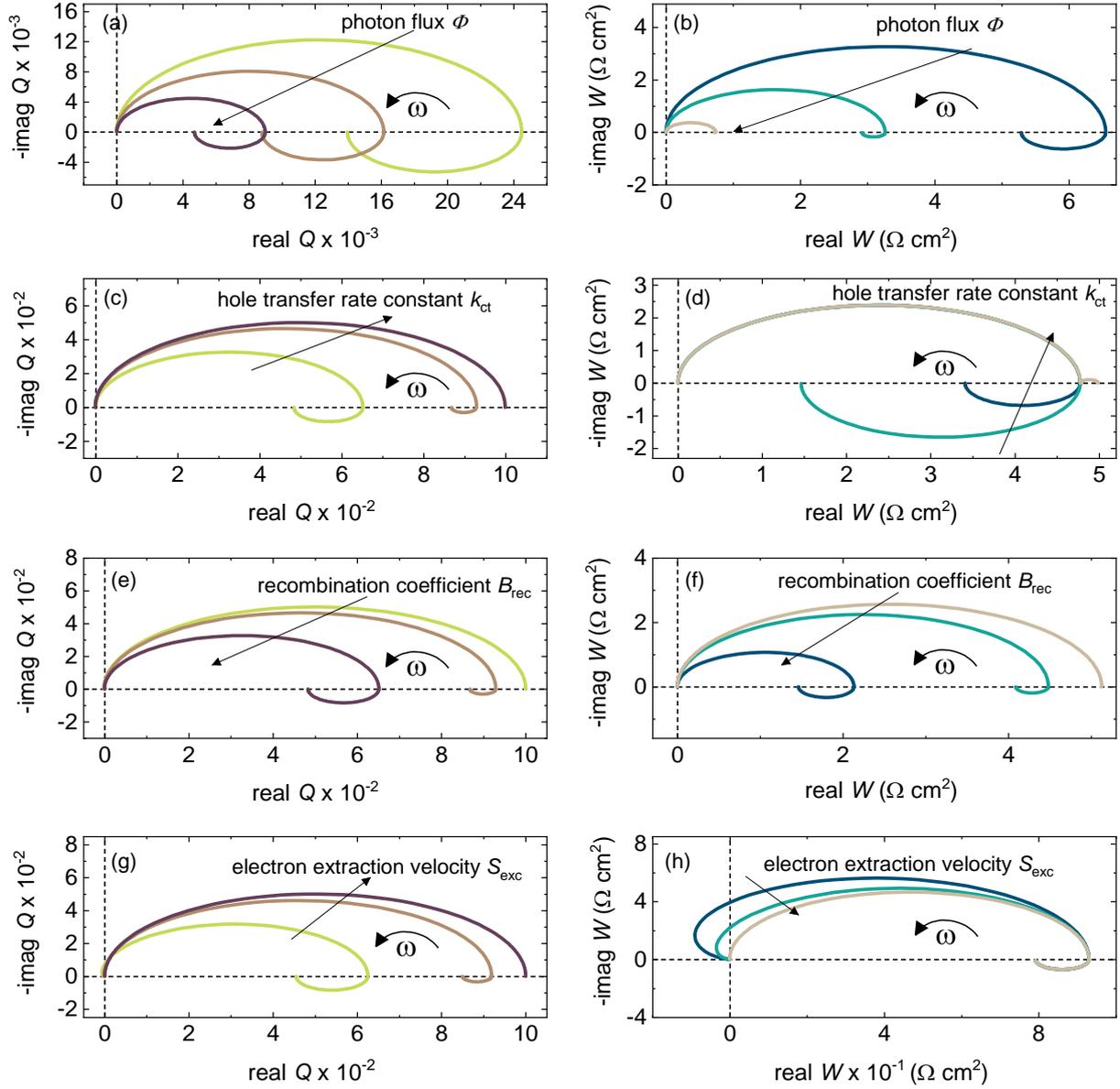

**Figure 8** Simulated (a, c, e, g) IMPS and (b, d, f, h) IMVS spectra using the developed analytical model, as a function of (a, b) DC photon flux $\Phi$, (c, d) hole transfer rate constant $k_{ct}$, (e, f) recombination coefficient $B_{rec}$ and (g, h) electron extraction velocity $S_{exc}$. In all cases except (c) and (d), slow hole transfer to the electrolyte is assumed (i.e. $\tau_{ct} \gg \tau_{exc}$). See table S2 in the SI for the list of simulation parameters. IMPS spectra were simulated at 1.23 V vs $V_{RHE}$ while IMVS spectra were simulated at open-circuit conditions (250 W/m² white LED illumination).

We now proceed to analyse the predicted evolution of the IMPS and IMVS spectra as a function of applied voltage from the analytical model, shown in figures 9(a) and 9(c). These spectra are simulated for the case of slow hole transfer to the electrolyte i.e. $\tau_{ct} \gg \tau_{exc}$, to be consistent with experimental observations. The IMPS spectra show the typical feature of two arcs - one in the upper quadrant at high frequencies and one in the lower quadrant at low frequencies. The IMVS spectra also show two arcs in the upper quadrant, with a small arc at high frequencies followed by a much larger arc at lower frequencies. The IMPS spectra in figure 9(a) show a trend of increasing HF arc width versus applied anodic voltage, with increasing magnitude of the HF and LF intercepts on the real axis. This can be understood by the dependence of the electron extraction velocity $S_{exc}$ on the electric field $F$ in equation 23. For large anodic voltages, there is a large electric field through the photoanode which results in $S_{exc}$ being proportional to the electric field ($S_{exc} \propto \mu_n F$). This means improved extraction of electrons at larger anodic voltages, which leads to the increased magnitudes of the HF and LF intercepts in the IMPS spectra seen in figure 9(a). The HF arc in the IMVS spectra does not evolve with anodic voltage, while the LF arc becomes larger with increasing anodic voltage.

These trends can be better understood by observing the evolution of the characteristic time constants, shown in figures 9(b) and 9(d) for IMPS and IMVS respectively. These time constants are typically obtained from the inverse maxima of the negative imaginary part of the transfer function versus angular frequency $\omega$ ($\omega = 2\pi f$, where $f$ is the measurement frequency) or equivalently, from the inverse angular frequency of the peak of the arc or arcs observed in the plot of real versus negative imaginary part of the transfer function (commonly referred to as a Nyquist plot). Recently, it has been shown that this method does not reveal all the time constants present in the data.[12a] For the case of non-ideal charge extraction at the collecting contact, a transition of the real part of the transfer function to negative values is observed, as also seen in figures 8(g) and 8(h). The time constant for charge extraction (equation 33) is embedded in this transition and cannot be extracted using the traditional analysis of the negative imaginary part of the transfer function. Simultaneous extraction of both the rise and decay time constants from the data has been achieved using a modified transfer function that focusses on the evolution of its real part versus $\omega$,[12a] which is discussed in detail in section A7 in the SI and applied to the simulations in figure 8 to determine $\tau_{exc}$. Therefore, we conclude that the spectra in figures 9(a) and 9(c) yield three time constants each – a fast time constant $\tau_1$ from the transition of the real part to negative values at high frequencies, followed by two time constants ($\tau_2$, $\tau_3$) that correspond to the two arcs observed in the spectra.

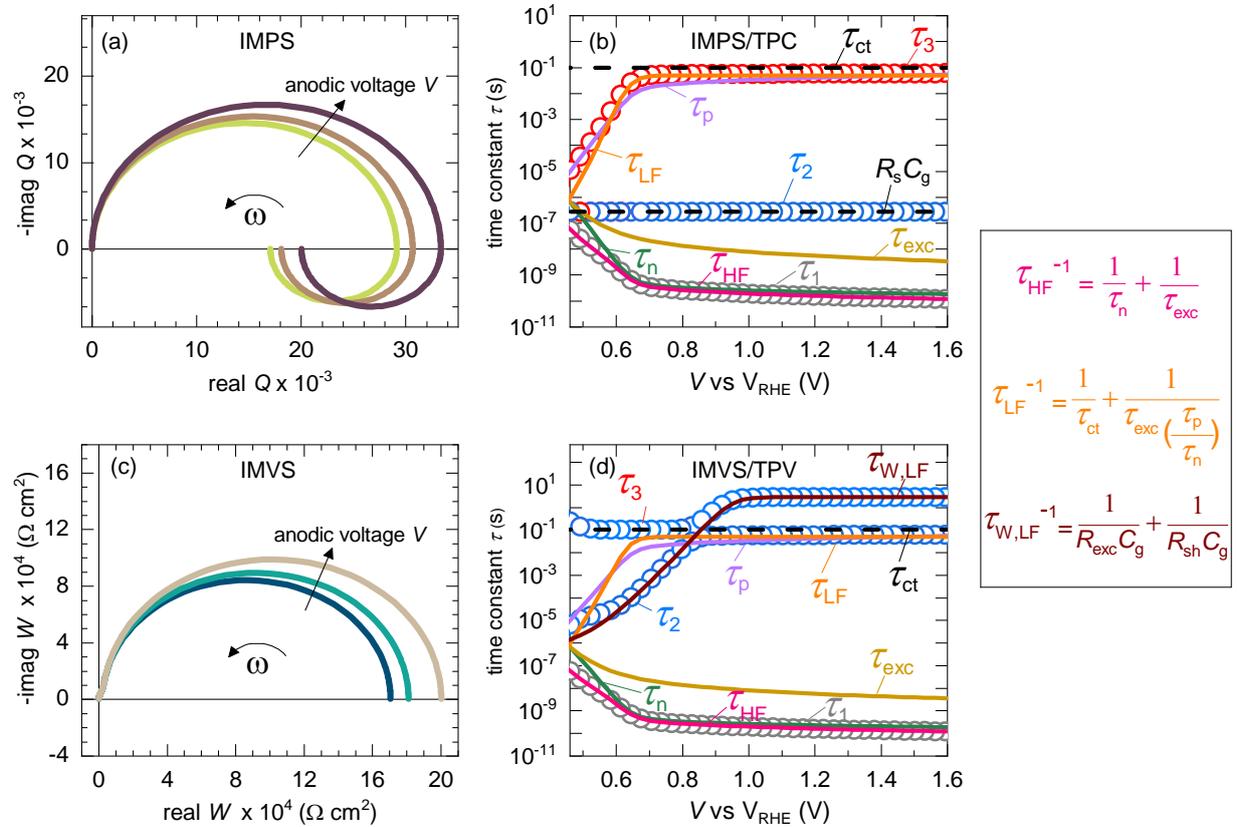

**Figure 9** Simulated (a) IMPS and (c) IMVS spectra using the developed model (slow hole transfer to the electrolyte i.e. $\tau_{ct} \gg \tau_{exc}$ is assumed), versus applied voltage. (b) and (d) show

the predicted effective time constants ($\tau_1$, $\tau_2$ and $\tau_3$, open markers) as a function of applied voltage. The box on the right-hand side shows the different time constants that comprise the observed effective time constants in (b) and (d). These time constants are also the same time constants that are predicted by the analytical model for the corresponding time domain techniques (i.e. IMPS→TPC and IMVS→TPV). We note that the time constants are the inverse eigenvalues of the matrices for IMPS and IMVS shown in section A6 in the SI. These eigenvalues can sometimes exchange with each other depending on their magnitudes, which has been corrected in (b) and (d) for ease of viewing. See table S2 in the SI for the list of simulation parameters. The hole transfer rate constant was set to $k_{ct} = 10$ s$^{-1}$ to maximise the visibility of the time constants in (b) and (d). This larger value of $k_{ct}$ (compared to the spectra in figure 8(d)) causes the low-frequency arc in the IMVS spectra to disappear.

We first focus on the high-frequency time constant $\tau_1$ obtained from the IMPS and IMVS spectra in figure 9. This time constant is determined by a parallel combination of the recombination lifetime for electrons $\tau_n$ (equation 34) and the electron extraction time constant $\tau_{exc}$ ($\tau_{HF}$ in figure 9(b) and 9(d)). In addition, $\tau_{HF}$ does not evolve significantly (i.e. over orders of magnitude) with applied anodic voltage. This is because of two factors – the first being the linear dependence of the electron extraction velocity $S_{exc}$ on the electric field ($S_{exc} \propto \mu_n F$, see equation 23) and hence applied voltage, and the second being the relatively voltage-independent magnitude of the hole concentration $p_s$ (see figure S4 in the SI). The latter effect is strongly influenced by the slow rate of hole transfer $k_{ct}$ to the electrolyte. However, at cathodic voltages, both $\tau_n$ and $\tau_{exc}$ (and hence $\tau_{HF}$) show a sharp increase in their magnitudes. The intermediate-frequency time constant $\tau_2$ in the case of IMPS is the voltage-independent product of the series resistance $R_s$ and the geometric capacitance $C_g$. This effect is generally referred to in the literature as *RC attenuation*.[6a, 29] In the case of IMVS, the intermediate-frequency time constant $\tau_2$ at large anodic voltages is a parallel combination of the hole transfer time constant $\tau_{ct}$, and $\tau_{exc}$ scaled by the ratio of the hole and electron recombination lifetimes. The slowest time constant $\tau_3$ for IMPS is the same as the intermediate-frequency IMVS time constant ($\tau_{LF}$ in figure 9(b) and 9(d)), while for IMVS, it is a parallel combination of attenuation via the exchange resistance $R_{exc}$ or the shunt resistance $R_{sh}$ ($\tau_{W,LF}$ in figure 9(d)). Moving towards cathodic voltages, in the case of IMVS, the exponential voltage-dependence of $R_{exc}$ (equation 31) makes $\tau_3$ and $\tau_2$ exchange with each other, leading to $\tau_3$ being dominated by $\tau_{ct}$ while $\tau_2$ is dominated by the $R_{exc}C_g$ product. A summary of the predicted time constants are shown in table 1.

**Table 1** Predicted high-frequency (HF), intermediate-frequency (IF) and low-frequency (LF) time constants using the developed analytical model for IMPS/TPC and IMVS/TPV measurements of a thin-film photoanode, in the limiting condition of slow hole transfer to the electrolyte.

| Experiment | HF time constant | IF time constant | LF time constant |
|---|---|---|---|
| IMPS/TPC | $\tau = \left(\dfrac{1}{\tau_n} + \dfrac{1}{\tau_{exc}}\right)^{-1}$<br><br>electron recombination in parallel with electron extraction | $\tau = R_s C_g$<br><br>RC attenuation | $\tau = \left(\dfrac{1}{\tau_{ct}} + \dfrac{1}{\tau_{exc}\left(\dfrac{\tau_p}{\tau_n}\right)}\right)^{-1}$<br><br>hole transfer to electrolyte, in parallel to electron extraction with a pre-factor |

| IMVS/TPV | $\tau = \left(\dfrac{1}{\tau_n} + \dfrac{1}{\tau_{exc}}\right)^{-1}$ | $\tau = \left(\dfrac{1}{\tau_{ct}} + \dfrac{1}{\tau_{exc}\left(\dfrac{\tau_p}{\tau_n}\right)}\right)^{-1}$ | $\tau = \left(\dfrac{1}{R_{exc}C_g} + \dfrac{1}{R_{sh}C_g}\right)^{-1}$ |
|---|---|---|---|
| | electron recombination in parallel with electron extraction | hole transfer to electrolyte, in parallel to electron extraction with a pre-factor | attenuation via the parallel combination of the electron exchange resistance and shunt resistance |

*Drift-diffusion simulations*

To test the validity of the predictions of our analytical model, we carry out full drift-diffusion simulations of the photoanode response (a basic overview of drift-diffusion simulations is provided in section A1 in the SI and corresponding parameters are shown in table S1 in the SI). We simulate the effect of different device parameters on the IMPS spectra at 1.23 V anodic voltage and IMVS spectra at open-circuit conditions, for the case of slow hole transfer to the electrolyte (i.e. $\tau_{ct} \gg \tau_{exc}$). Figures 10(a) and 10(b) show the effect of the DC light intensity on the IMPS and IMVS spectra respectively. The IMPS spectra show a HF arc in the upper quadrant and a LF arc in the lower quadrant, whose intercepts on the real axis become smaller with increasing light intensity. The IMVS spectra show two arcs in the upper quadrant, whose widths decrease with increasing light intensity. These trends are identical to that predicted by our analytical model in figures 8(a) and 8(b) – due to the slow hole transfer, photogenerated electron-hole pairs due to increasing light intensity increase the recombination rate, leading to a reduction in the magnitudes of the arc widths and intercepts versus increasing light intensity. However, in the case of IMVS, the analytical model predicts the second arc to exist in the lower quadrant (figure 8(b)), while the drift-diffusion simulations show the second arc in the upper quadrant. We note that our analytical model predicts that the arc in the lower quadrant is shifted to the upper quadrant for large $k_{ct}$ values (see figure 8(d)).

The evolution of the IMPS spectra as a function of the hole transfer rate constant $k_{ct}$ and the recombination coefficient $B_{rec}$ are shown in figures 10(c) and 10(e) respectively. We observe an arc in the upper quadrant at high frequencies followed by an arc in the lower quadrant at low frequencies. The magnitudes of the intercepts of the HF and LF arcs on the real axis become larger (with the width of the LF arc becoming smaller) for higher $k_{ct}$ values due to an increased photocurrent arising from efficient hole transfer to the electrolyte. Conversely, increasing $B_{rec}$ values reduce the magnitudes of the intercepts. These trends are also consistent with the predictions of the analytical model (figures 8(c) and 8(e)). In the case of IMVS, larger $k_{ct}$ values (figure 10(d)) lead to a larger magnitude of the arcs' intercepts on the real axis, while increasing $B_{rec}$ values (figure 10(f)) cause a reduction in the magnitude of these intercepts. Such a trend can be understood by the fact that smaller values of $k_{ct}$ and larger values of $B_{rec}$ lead to an increased recombination rate, which implies a smaller recombination resistance. These trends are also predicted by the analytical model (figure 8(d) and 8(f)). Finally, the effect of the electron extraction velocity $S_{exc}$ on the IMPS and IMVS spectra is shown in figures 10(g) and 10(h) respectively. Increasing $S_{exc}$ values have a similar effect to that of increasing $k_{ct}$ values, where for IMPS, the HF and LF arc intercepts on the real axis increase in magnitude. In case of IMVS, the two HF arcs increase in width for increasing $S_{exc}$ values. Furthermore, the influence of $S_{exc}$ can be seen in the high-frequency region of the spectra in figures 10(g) and 10(h), where the real part of the transfer function shows a pronounced transition to negative values before reaching the origin. As discussed in the last section, this region of the spectra allows determination of the electron extraction time constant using a suitable data transformation, explained in section A7 in the SI. Finally, drift-diffusion simulations of the IMPS and IMVS spectra versus applied voltage are shown in figure S7 in the SI, which also show a good agreement with the trends predicted by the analytical model in figure 9. In summary, we conclude that our analytical model consistently captures the trends of the IMPS and IMVS spectra as a function of different DC parameters, which is confirmed using full drift-diffusion simulations.

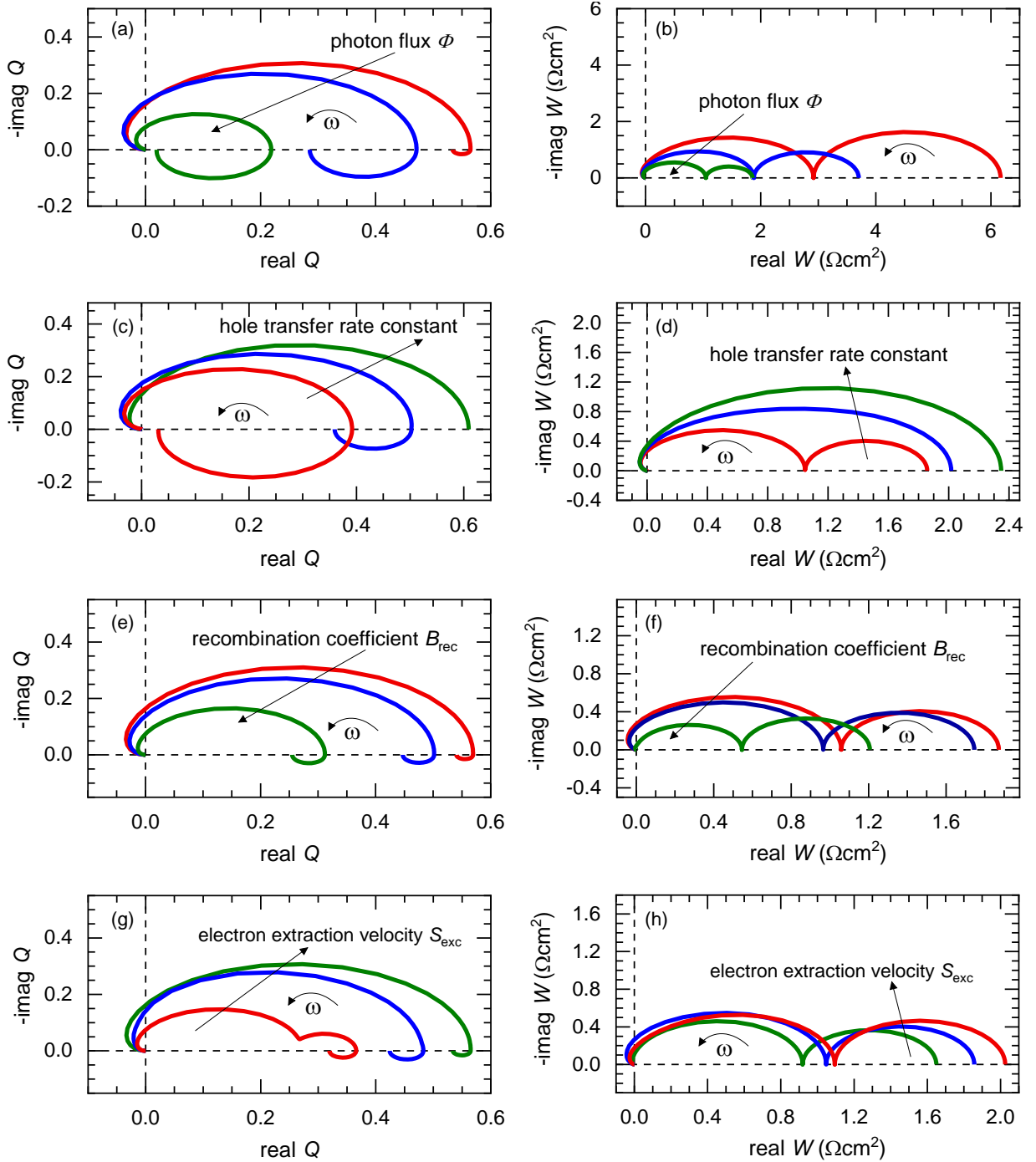

**Figure 10** Drift-diffusion simulations of (a, c, e, g) IMPS (at 1.23 V anodic voltage) and (b, d, f, h) IMVS (at open-circuit, 25 mW/cm² white LED illumination) spectra. Simulation parameters are shown in table S1 in the SI, with the doping density $N_d = 0$ cm$^{-3}$, mobilities of electrons and holes in the photoanode/electrolyte interface layer set to $\mu = 5 \times 10^{-12}$ cm$^2$V$^{-1}$s$^{-1}$. These simulations are made as a function of (a, b) DC photon flux $\Phi$, (c, d) hole transfer rate

constant by variation of the interface layer mobility, (e, f) recombination coefficient $B_{\text{rec}}$ and (g, h) electron extraction velocity $S_{\text{exc}}$. In all cases where the hole transfer rate is not varied, slow hole transfer to the electrolyte is assumed (i.e. $\tau_{\text{ct}} \gg \tau_{\text{exc}}$).

*Analysis of experimental spectra*

We now proceed to characterize our hematite photoanode (current-voltage curves shown in figure 5) using a combination of IMPS, IMVS and TPV measurements, followed by interpretation of the spectra using the analytical model. We clarify that this model has been developed to serve as a general, base model that captures the fundamental physics of charge generation, recombination, electron extraction and interfacial hole transfer in a consistent manner, for both the steady-state and transient response in a semiconductor photoanode. It therefore does not capture the physics of specific effects occurring in different systems, for example, charge accumulation at surface states,[9c] formation of surface-bound intermediates[6b] and Fermi-level pinning.[30] Inclusion of specific effects such as trapping-de-trapping of charges via surface states and the effect of the potential drop in the Helmholtz layer is beyond the scope of this work. We therefore use the model to determine only the rate constants or characteristic time constants of hole extraction at the photoanode/electrolyte interface and electron extraction at the collecting electrode, since these parameters should not be affected significantly by the specific nature of the interfacial recombination in our hematite photoanodes.

The IMPS and IMVS measurements were made across a range of bias voltages under 25 mW/cm² white LED illumination within a frequency range of 100 mHz - 20 kHz, shown in figures 11(a) and 11(b) respectively. The IMPS spectra show two arcs, one in the upper quadrant at high frequencies and one in the lower quadrant at low frequencies. For larger anodic voltages, both the HF and LF intercepts on the real axis (that correspond to the arcs in the upper and lower quadrant respectively) increase in magnitude. This behaviour is consistent with the predictions from both the analytical model (figure 9(a)) and the drift-diffusion simulations (figure S7(a) in the SI), and occurs due to improved electron extraction to the collecting contact arising from a larger electric field in the bulk. We note that the arc in the upper quadrant is incomplete due to the upper limit (20 kHz) of the frequency range of our measurement setup. In the case of IMVS, we observe a single arc in the upper quadrant whose width becomes larger for increasing anodic voltages. The IMVS arcs are incomplete due to the chosen lowest frequency of the measurement (100 mHz), which confirms the very slow response of the hematite photoanodes. Since the IMPS and IMVS measurements capture the LF response of the device but not the HF response, we carried out TPV measurements to determine the rise time constants. An example of a TPV spectrum of our device is shown in figure 11(c), which was fitted using equation 4 assuming $\tau_{\text{rise}} \ll \tau_{\text{decay}}$ (as confirmed by the full set of TPV spectra shown in figure S8 in the SI) within the measured time range, to calculate the rise time constant. Figure S9 in the SI shows the corresponding fits of the entire set of TPV spectra.

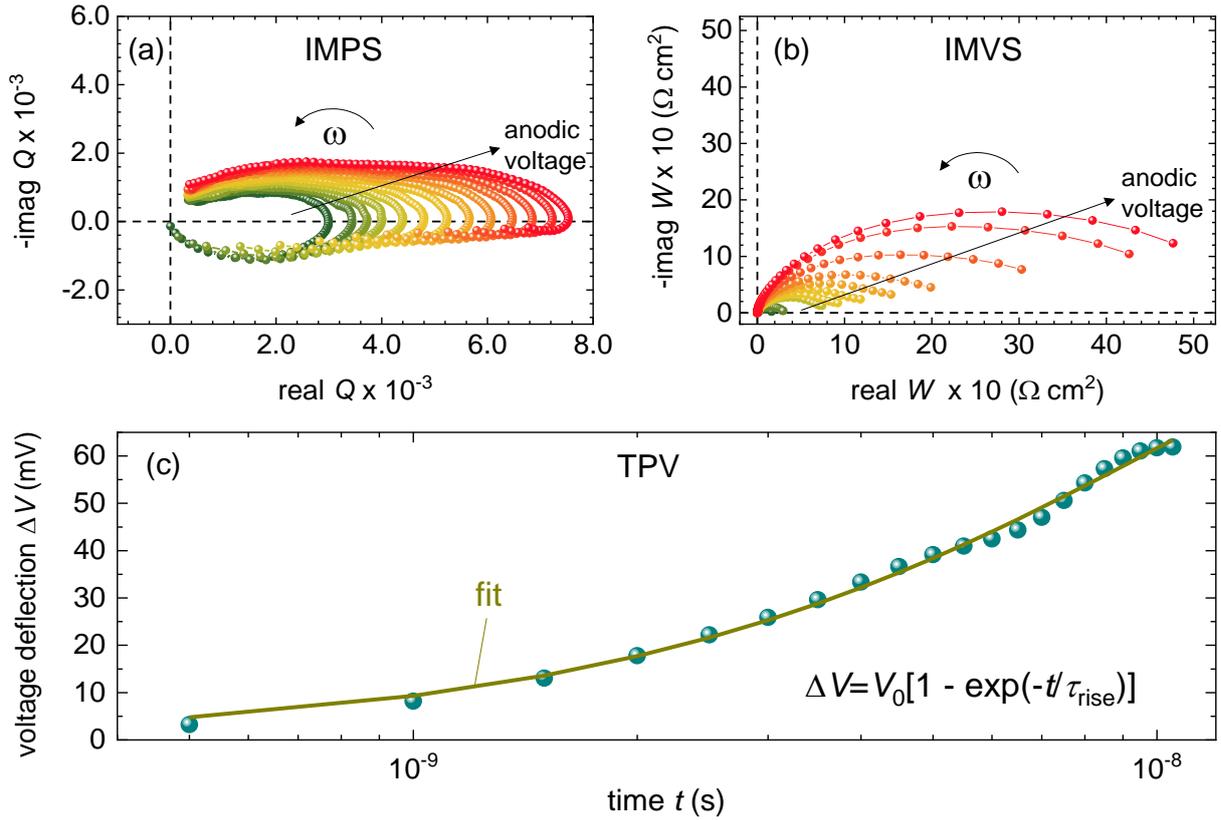

**Figure 11** Measured (a) IMPS and (b) IMVS spectra of a hematite photoanode in KOH 1M solution, as a function of applied anodic voltage under 25 mW/cm² white LED illumination. (c) Example of the measured TPV rise spectra and fit using equation 4 (see inset) to determine the rise time constant. For ease of view, the decay of the photovoltage deflection is not shown. See figure S8 in the SI for the set of full (rise and decay) TPV spectra and figure S9 in the SI for the fits of the photovoltage deflection rise to obtain the rise time constants.

Figure 12 shows the calculated time constants from the IMPS, IMVS and TPV measurements. IMPS yields two time constants – a voltage-independent IF time constant with a magnitude of $10^{-4}$ s and a LF time constant with a magnitude ranging between $0.08 - 0.5$ s. The IMVS measurements yield a LF time constant that is very similar in magnitude to the IMPS LF time constant. In case of the TPV measurements, we obtained a voltage-independent rise time constant with a magnitude of $\cong 10^{-8}$ s (figure 12(b)). Using our analytical model, we can assign these time constants to different physical mechanisms occurring in the device. Figure 9(d) shows that the slowest time constant in the case of IMVS corresponds to the time constant of hole transfer to the electrolyte $\tau_{ct}$, until the attenuation via the parallel combination of the exchange resistance and shunt resistance (see table 1) takes over at high anodic voltages. We thus assign the IMVS time constant at the most cathodic voltage value as $\tau_{ct}$, obtaining a $k_{ct}$ value of 16 s⁻¹. In the case of the IMPS time constants, the IF time constant corresponds to $RC$ attenuation (i.e. $\tau_{IF,IMPS} = R_s C_g$, confirmed by the IS data in figure S10 in the SI) and is hence not useful for our analysis. The LF time constant in IMPS is again strongly dominated by $\tau_{ct}$ according to figure 9(b) and hence does not provide any additional useful information. From figure 9(d) and table 1, we observe that the TPV rise time constant is a parallel combination of the electron extraction time constant $\tau_{exc}$ and the electron recombination lifetime $\tau_n$. Figure 9(d) further shows that upon moving towards cathodic voltages, $\tau_n$ becomes larger in magnitude and approaches $\tau_{exc}$, eventually becoming larger than $\tau_{exc}$ for higher cathodic voltages. Therefore, we assume that the HF time constant, which corresponds to the rise time constant in TPV (measured at the light open-circuit potential, which corresponds to cathodic voltages), is dominated by $\tau_{\text{exc}}$. Assuming $d_{\text{int}} = 10$ nm and setting $\tau_{\text{exc}} = 10^{-8}$ s, we can calculate the electron exchange velocity $S_{\text{exc}}$ using equation 33. We thus obtain a value of $S_{\text{exc}} = 100$ cm/s. Furthermore, at open-circuit conditions close to or above the 1 sun open-circuit voltage (which is assumed to be the flatband voltage), we can Taylor-expand the exponential term in equation 23 and replace the absorber layer thickness $d$ ($\cong 100$ nm for our hematite photoanodes) with an effective thickness $d_{\text{eff}} = d - d_{\text{int}}$ across which the electrons are transported to the collecting contact, obtaining

$$S_{\text{exc}}(V_{\text{oc}}(1\text{ sun})) \cong \frac{\mu_n F\left(1 - \frac{qd_{\text{eff}}F}{k_B T}\right)}{\exp(-1) - \left(1 - \frac{qd_{\text{eff}}F}{k_B T}\right)} = \left[\frac{1}{1 - \exp(-1)}\right]\left(\frac{k_B T}{q}\right)\frac{\mu_n}{d_{\text{eff}}}. \tag{41}$$

Using equation 41, we estimate a value of $\mu_n = 0.022$ cm$^2$V$^{-1}$s$^{-1}$ for the electron mobility in the hematite layer, which is comparable to the values reported in literature for similar hematite configurations.[31]

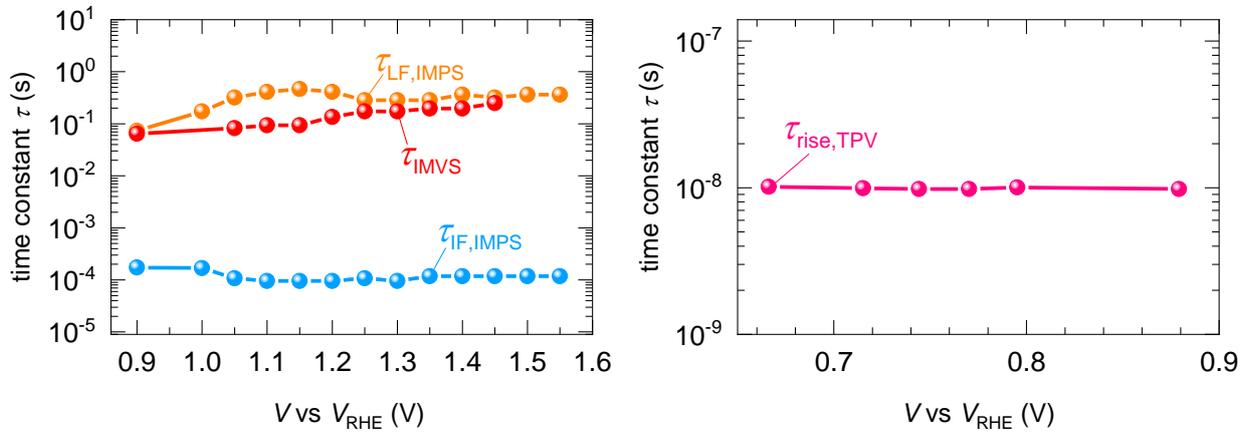

**Figure 12** Calculated (a) IMPS and IMVS time constants from the experimental spectra shown in figure 11(a, b) and (b) rise time constants from the TPV measurements shown in figure S8 in the SI (fits shown in figure S9 in the SI).

**Conclusions**

Further development in the efficiency of operation of photoanodes requires a fundamental understanding of the physical mechanisms occurring within them - in particular, the charge carrier transport, recombination and transfer at the photoanode/electrolyte interface or the collecting electrode. While existing optoelectronic models capture the effect of specific mechanisms occurring in bulk or at the photoanode/electrolyte interface, they cannot be applied to consistently reproduce both the steady-state and transient optoelectronic response (time and frequency domain) of the photoanode. Furthermore, these models do not account for a fundamental feature in all photovoltaic devices - the difference between the internal (average quasi-Fermi level splitting within the photoanode) and external voltage that drives the photocurrent and the excess recombination arising from non-ideal electron extraction at the collecting electrode. Finally, these models assume a highly doped photoanode with a thin depletion layer at the photoanode/electrolyte interface, which has not been confirmed experimentally across different photoanode materials.

To solve these problems, we have developed an analytical model that assumes that the potential distribution within the photoanode is determined by the charges on the electrode, rather than a doping density in the bulk. In addition to the processes of charge generation, recombination and slow hole transfer at the photoanode/electrolyte interface, this model explicitly accounts for the dependence of the electron current on the difference between the quasi-Fermi level splitting within the photoanode under illumination and the applied external voltage. Furthermore, it also accounts for the effect of non-ideal electron extraction at the collecting electrode and its influence on the recombination rate at the photoanode/electrolyte interface. The model is able to consistently simulate both the steady-state and transient response of the semiconductor photoanode, allowing comparison and analysis of different light and voltage-modulated techniques both in the time domain (e.g. transient photovoltage (TPV), transient photocurrent (TPC)) and frequency domain (e.g. intensity-modulated photocurrent (IMPS) and intensity-modulated photovoltage spectroscopy (IMVS)), to obtain a unified analysis and verification of the extracted parameters from the data. We verified the model's predicted IMPS and IMVS spectra across a range of DC parameters using full drift-diffusion simulations, obtaining a good agreement with each other. Our model can thus serve as a base model for modelling the optoelectronic response of photoanodes, which can be modified to add specific effects that are relevant for the chosen materials or interfaces under investigation.

We used the model to interpret the measured time constants from small-perturbation measurements of hematite photoanodes both in the frequency domain (IMPS and IMVS) and time domain (TPV). We calculated a hole transfer rate constant of $k_{ct} = 16$ s$^{-1}$, at the photoanode/electrolyte interface, confirming the well-known slow kinetics of holes in this region. We also determined an electron extraction velocity of 100 cm/s, that corresponds to an electron mobility of 0.022 cm$^2$V$^{-1}$s$^{-1}$ in the hematite layer. Furthermore, the dependence of the electron extraction on the electric field within the photoanode causes the model to naturally reproduce the linear dependence of the photocurrent on the applied voltage that is commonly observed in hematite photoanodes.

**Acknowledgements**
This work has been partially funded by the Project 'Network 4 Energy Sustainable Transition – NEST' (code PE0000021) under the National Recovery and Resilience Plan (NRRP) (European Union – NextGenerationEU and Ministero dell'Università e della Ricerca-MUR). S.R. acknowledges that this work is funded by the Deutsche Forschungsgemeinschaft (DFG, German Research Foundation) – project number 539945054. S.R. and T.K. acknowledge support from the Helmholtz Association via the programme-oriented funding (POF IV) and SolarTap project. Open access publication funded by the DFG – 491111487.

**Author contributions**
Conceptualization: S.R., T.K.
Methodology: S.R., P.R.
Sample preparation: P.R., B.K.
Investigation: P.R., S.R.
Visualization: P.R., S.R., B.K., C.M., T.K.
Supervision: S.R., C.M.
Writing – original draft: P.R., S.R.
Writing – review and editing: P.R., S.R., B.K., C.M., T.K.

**Data availability**
All simulation files are uploaded to the Zenodo database with the identifier 10.5281/zenodo.14872104.

**Conflicts of interest**

There are no conflicts of interest to declare.